# Comparison of dynamic mechanical properties of non-superheated and superheated A357 alloys

M. N. Mazlee*, J. B. Shamsul and H. Kamarudin
School of Materials Engineering, Universiti Malaysia Perlis (UniMAP)
Kompleks Pusat Pengajian Jejawi 2, Taman Muhibah, Jejawi
02600 Arau, Perlis, Malaysia
mazlee@unimap.edu.my

**Abstract**

The influence of superheat treatment on the microstructure and dynamic mechanical properties of A357 alloys has been investigated. The study of microstructure was performed by optical microscope. Dynamic mechanical properties (storage modulus, loss modulus and damping capacity) were measured by the dynamic mechanical analyzer (DMA). Microstructure showed coarser and angular eutectic Si particles with larger α-Al dendrites in non-superheated A357 alloy. In contrast, finer and rounded eutectic Si particles together with smaller and preferred oriented α-Al dendrites have been observed in superheated A357 alloy. Dynamic mechanical properties showed an increasing trend of loss modulus and damping capacity meanwhile a decreasing trend of storage modulus at elevated temperatures for superheated and non-superheated A357 alloys. The high damping capacity of superheated A357 has been ascribed to the grain boundary damping at elevated temperatures.

## 1. Introduction

In the last few decades, Al-Si-Mg alloy is one of the most commercially alloys used in the automotive and aircraft industries which attributed by its excellent combination of good castability and mechanical properties as well as good corrosion resistance and weldability [1-3]. Solidification process plays a pivotal role in determining the resulted physical and mechanical properties of the alloys. The solidification sequence of alloy Al-Si-Mg alloy consists mainly of three phase transformations by referring to the cooling curves namely the formation of aluminum dendrites (α-Al dendrites), the main binary eutectic reaction and the formation of ternary and/or quaternary eutectic phases such as $Mg_2Si$ and/or Fe-bearing intermetallics [4].

Melt superheat treatment was the alternative non-chemical and thermal treatment that has been practically utilized for grain refinement to the molten metal in aluminum alloys [5, 6] and magnesium alloys [7, 8]. Superheat treatment combined with grain refiner in production of Al-Si alloys has been proven towards improving mechanical properties [6, 9, 10]. Grain refinement by the addition of grain refiner associated with increased amount of α-Al dendrites has contributed to the increasing of damping capacity [3, 11].

## 1.1 Dynamic mechanical properties

Damping capacity (tan ϕ) is a measure of a material's ability to dissipate elastic strain energy during mechanical vibration or wave propagation [12, 13]. The damping capacity of structural material has been obligated to recognize as one of significant factor for material selection [14].

In an ideally elastic materials, $\phi = 0$ and $\sigma/\varepsilon = E$, the Young's modulus. However, most materials are anelastic owing to some energy dissipation which occurs in the same frequency range as the imposed stress, therefore the strain lags behind the stress and ϕ is not zero. The ratio of σ/ε is the complex modulus, $E^*$ and is defined as

$$E^* = E' + iE'' \qquad (1)$$

where $E'$ and $E''$ are the components of $E^*$ and known as the storage modulus and loss modulus respectively. The ratio of the two elastic moduli $E''/E' = \tan \phi$ is the damping capacity of the material. This term is dimensionless and shows the ratio of energy lost (dissipated by heat) per cycle to energy stored [15].

Only few studies are conducted on damping of Al-Si based alloys. Yijie *et al.* [3, 11] found the increment of damping capacity which attributed by the grain refinement with the addition of nano grain refiner in A356 alloy. According to Zhou *et al.* [14], the improvement of damping was linked with dislocation movements induced by the thermal stresses due to the coefficient of thermal expansion mismatch between the Al matrix and Si particles in Al-11.8 Si alloy. Meanwhile Lee [16] reported that the maximum specific damping capacity was coincided with the maximum age hardening condition but simultaneously decrease with over aging in Al-7 Si-0.3 Mg alloy with T6 treatment. This paper presents the results of an investigation of the effects of melt superheat treatment on the microstructure and dynamic mechanical properties of A357 alloys.

## 2. Experimental

The raw material used as a matrix alloy in this research work was primary cast ingot Al-Si-Mg alloy supplied by National Centre for Machinery & Tooling Technology (NCMTT), SIRIM Berhad, Malaysia. The alloy has been cast by continuous casting process and was delivered in the form of bar. The chemical composition of cast alloy was analysed using arc spark spectroscopy (SPECTROMAXx, Germany) and the composition was complied with A357 alloy. Table 1 shows the chemical composition of primary cast ingot A357 alloy.

The primary cast ingot A357 alloy was melted in crucibles by using electric furnace. Superheated cast A357 alloy was prepared by heating up the primary cast ingot Al-Si-Mg alloy to the melt superheat temperature of 900°C for 1 hour before poured into the preheated stainless steel mould via conventional gravity casting technique. Meanwhile, the non-superheated cast A357 was prepared by melting the primary cast ingot Al-Si-Mg alloy at 750ºC for 1 hour before poured into the same mould.

Table 1    Chemical compositions of primary cast ingot A357 alloy.

| Elements | Si | Mg | Fe | Mn | Ti | Al |
|---|---|---|---|---|---|---|
| Composition (wt. %) | 7.24 | 0.54 | 0.12 | <0.10 | 0.10 | Balance |

The microstructures of all the specimens were characterized by light microscope. Specimens were prepared by the standard metallography methods of cutting and mounting followed by wet grinding on a series of SiC papers. Finally, the specimens were polished with 6 µm, 3 µm and 1 µm diamond suspension using napless cloth. The etchant used was 0.5% HF in order to reveal the morphologies of α-Al dendrites and eutectic Si particles in the microstructures.

A dynamic mechanical analyzer (Pyris Diamond DMA model, USA) was used to measure storage modulus, loss modulus and damping capacity. Dynamic mechanical analysis was carried out in the three point bending mode using a dual cantilever system. The composite specimens were prepared in the form of rectangular bars with dimensions of 50 x 10 x 1.3 mm. The tested specimens were run at 2°C/minute heating rate from 30 to 400ºC with 100 µm strain at 10 Hz in a flowing purified nitrogen gas.

Differential scanning calorimetry (DSC) studies were carried out on the specimens of non-superheated and superheated as-cast A357 alloy specimens using DSC Q10 TA Instrument model. Parallel slices of 0.5 mm thick were cut from the specimens using precision diamond blade and discs about 5 mm in diameter were then punched from the slices. The specimens were tested at 10°C/min heating rate from 50 to 450ºC.

### 3.    Results and discussion

#### 3.1.  Microstructure

Figs. 1 and 2 show the microstructure of non-superheated and superheated A357 alloys respectively. Generally, both as-cast alloy microstructures showed the presence of α-Al dendrites and eutectic Si particles. Fig. 1 (a) shows the existence of coarser and angular eutectic Si particles surrounded with larger α-Al dendrites in non-superheated A357 alloy. The irregular and non-equiaxed α-Al dendrites of about 150 µm size were also observed in this alloy. These features were attributed to the slow solidification rate during casting. Fig. 1 (b) illustrates the coarse interdendritic of eutectic Si particles of about more than 10 µm size which segregated within the α-Al dendrites structure.

In Fig. 2 (a), a mixture of finer of about 100 µm size and rounded eutectic Si particles and also preferred oriented α-Al dendrites are observed in superheated A357 alloy. This is believed that the melt superheat prior to casting has influenced the microstructure. A high superheat treatment temperature of 900ºC has also assisted the degeneration of Si atom clusters. It was found also that superheat treatment in the current research tends to refine the eutectic Si particles and change the shape of α-Al dendrite to finer, rounded and preferred oriented α-Al dendritic structures.

The finer interdendritic of eutectic Si particles of about less than 10 µm are found to be segregated and agglomerated within the α-Al dendrites as shown in Fig. 2 (b). According to Wanqi *et al.* [17], if the superheat temperature is kept below 800ºC, there is

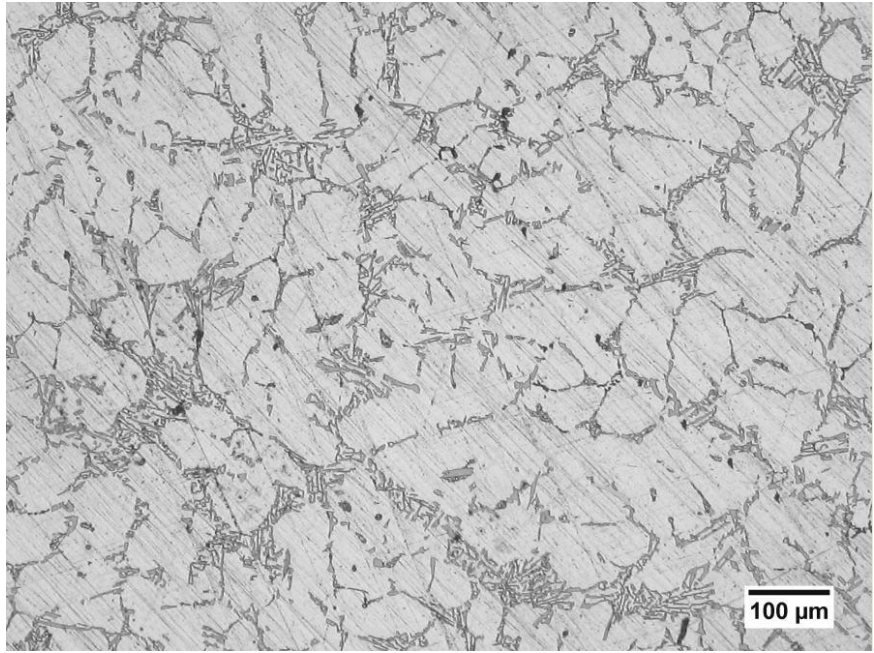

Fig. 1 (a) Optical microstructure of non-superheated A357 alloy at low magnification.

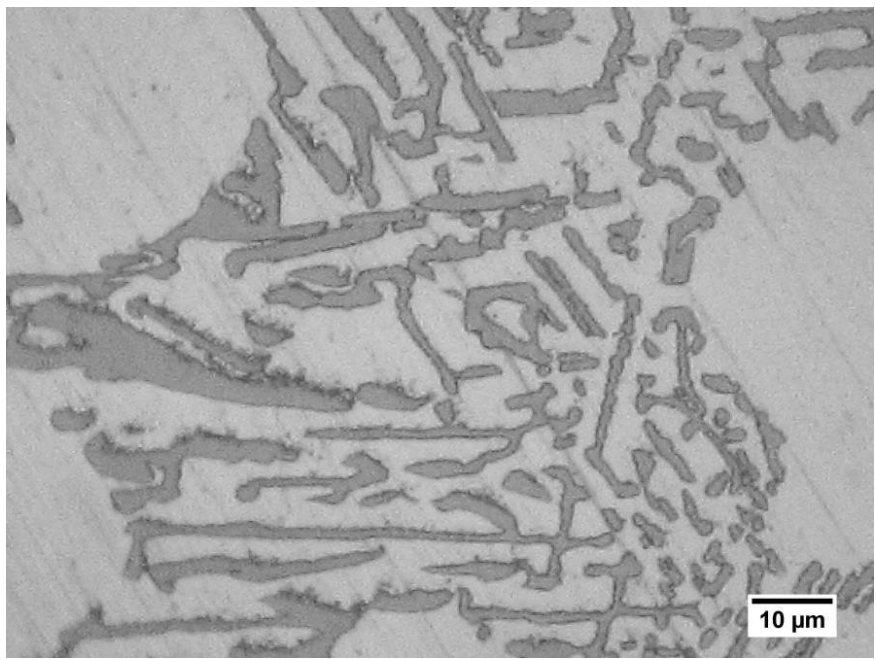

Fig. 1 (b) Optical microstructure of non-superheated A357 alloy at high magnification.

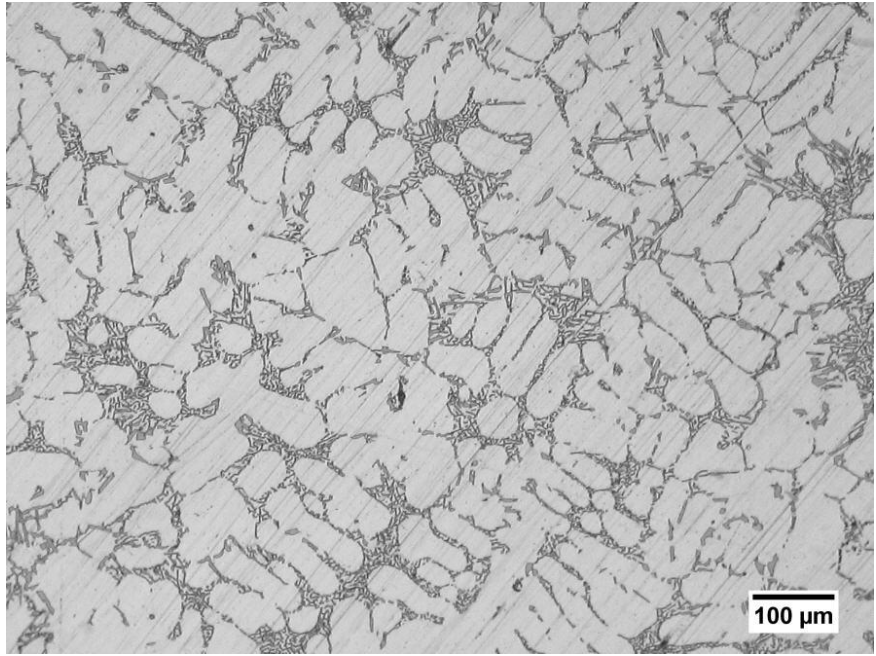

Fig. 2 (a) Optical microstructure of superheated A357 alloy at low magnification.

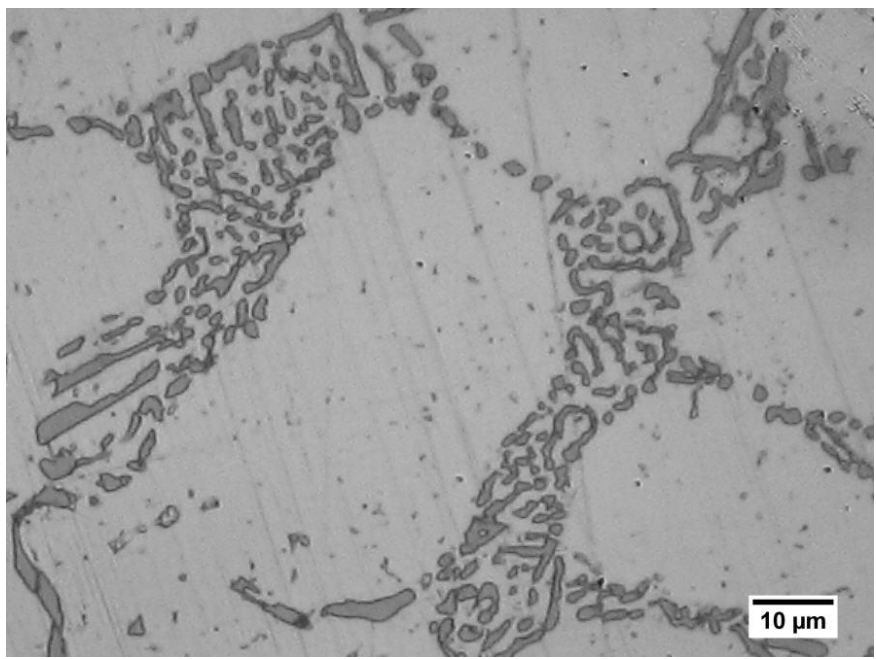

Fig. 2 (b) Optical microstructure of superheated A357 alloy at high magnification.

no obvious effect on the refinement of the Si particles in A356 alloy. However, over 800ºC, the Si particles are significantly refined. At melt temperatures of 900ºC, the refinement of the Si particles by melt superheat is nearly comparable to that obtained with 0.01 percent Sr addition. Wanqi *et al.* [17] contributed this refinement effect of superheat to the existence of Mg in the A356 alloy as they did not find any such effect in Al-Si binary alloys which containing no Mg.

### 3.2. Dynamic mechanical properties

Figs. 3, 4 and 5 show the dynamic mechanical properties behavior of non-superheated and superheated A357 alloy in terms of storage modulus, loss modulus and damping capacity at 10 Hz frequencies respectively. They show a general phenomenon of progressively decreasing storage modulus with increasing temperature (Fig. 3). In contrast, the loss moduli and damping capacity generally increase with increasing temperature (Figs. 4 and 5). It can be seen that the storage moduli, loss moduli and damping capacity are sensitive to the processing method of the alloys and testing temperature range.

Fig. 3 shows that the superheated A357 alloy retains a higher storage modulus compared to non-superheated A357 alloy at all temperatures. The storage modulus of superheated A357 alloy specimen is 50.08 GPa at 50ºC, decreases to 47.48 GPa at 200ºC followed by small increases to 47.77 GPa at 234ºC and reduce significantly to 43.28 GPa at 380ºC respectively. Meanwhile, the storage modulus of non-superheated A357 alloy specimen is 37.52 GPa at 50ºC, decreases to 34.90 GPa at 207ºC followed by small increases to 35.12 GPa at 234ºC before reduce significantly to 32.59 GPa at 380ºC respectively. Generally, both alloys show the same trend of smoothly decreasing values of storage modulus with increasing temperature with a rapid decrease at higher temperatures.

The higher storage moduli in superheated A357 alloy is attributed to the combined effect of finer and rounded interdendritic of eutectic Si particle together with smaller and preferred oriented α-Al dendrites compared to coarser and angular interdendritic of eutectic Si particle and larger α-Al dendrites in non-superheated A357 alloy. Preferred oriented α-Al dendrites have a profound effect to influence the enhancement of storage modulus as reported by textured structure [15].

The eutectic Si particles morphology has been found to play a vital role in determining the mechanical properties. Particle size, shape and spacing are factors that characterize Si morphology. A prolonged solution treatment on a cast Al-Si-Mg alloy has produced the spheroidized Si particles that contributed to the better tensile strength and elongation [18]. Ogris *et al*. [19] reported that eutectic silicon spheroidization treatment results in outstanding fracture elongation values (up to 18 percent) at good yield strength level (~230 MPa) in Al-Si alloys.

The loss modulus generally exhibited an increase at elevated temperatures. From Fig. 4, it can be seen that superheated A357 alloy specimen showed higher loss moduli than non-superheated A357 alloy specimen after 121ºC. The loss modulus of superheated A357 alloy is 1.09 GPa at 50ºC, overlaps about 1.17 GPa at 100ºC, start deviates about 1.20 GPa at 121ºC, increases to 1.77 GPa at 245ºC and to rise substantially to 2.52 GPa at 380ºC respectively. Meanwhile, the loss modulus of non-superheated A357 alloy specimen is 1.09 GPa at 50ºC, overlaps about 1.17 GPa at 100ºC, start deviates about 1.20 GPa at 121ºC, increases to 1.35 GPa at 229ºC and to rise substantially to 2.08 GPa at 380ºC respectively. Generally, loss modulus in superheated A357 alloy demonstrates a slowly increasing trend at below 210ºC followed by substantial increases within 210 to 380ºC. In contrast, loss modulus in non-superheated A357 alloy shows a decline at below 229ºC followed by considerable steady increases within 229 to 380ºC.

The loss moduli trend of superheated A357 alloy specimen in Fig. 4 has shown more efficient of energy loss relatively due to finer and rounded interdendritic of eutectic Si particles and combined structure of smaller and preferred oriented α-Al dendrites. The refinement of eutectic Si particles along with smaller α-Al dendrites play a partial role on the dissipation of elastic strain energy have been reported by Zhang *et al.* [3].

The damping capacity also generally exhibited an increase at elevated temperatures. The damping capacity of non-superheated and superheated A357 alloys exhibit an increase at elevated temperatures as can be seen from Fig. 5. The damping capacity of superheated A357 alloy specimen is 0.024 GPa at 50ºC followed by slowly increases to 0.039 GPa at 250ºC and lastly surges to 0.084 GPa at 380ºC respectively. Meanwhile, the damping capacity of non-superheated A357 alloy specimen is 0.032 GPa at 50ºC followed by slowly increases to 0.039 GPa at 250ºC and lastly surges to 0.064 GPa at 380ºC respectively. In overall, damping capacity of superheated A357 alloy is lower than non-superheated A357 alloy at below 190ºC but remain the same during overlapping of damping capacity trend line between 190 to 277ºC. At above 277ºC, superheated A357 alloy starts deviate and surges higher damping capacity than non-superheated A357 alloy until reach 380ºC.

It has been reported that, at above 400ºC, the damping capacity will reach the maximum peak. A prior work by Elomari *et al.* [20] has led to the conclusion that the mechanism responsible for the damping peak is probably related to the transformation processes (growth of precipitate) in the material. Fig. 6 has demonstrated the differential scanning calorimetry (DSC) thermographs of superheated and non-superheated A357 alloys at elevated temperatures respectively. It can be noted that damping capacity trend line of both alloys has overlapped between 190 to 277ºC. The overlap of damping capacity trend line has resembles the obvious area changes under DSC thermographs in Fig. 6.

Unequiaxed structure of α-Al dendrites in non-superheated and superheated A357 alloys respectively has contributed to the uneven trend lines of damping capacity as illustrated in Fig. 5. This finding was dissimilar to the smooth trend line of equiaxed grain has been observed by Zhang *et al.* [21].

### 3.3. Damping mechanisms

#### 3.3.1. Dislocation damping

Dislocation damping is noteworthy because it plays a critical role, not only in the damping response of crystalline material but also in the overall mechanical behavior of the materials [21]. According to Granato-Lücke theory [22, 23], the dislocation structure is assumed to consist of segments of length $L_N$ along which weak pinning points are distributed randomly, the dislocations are pinned by the fine precipitation.

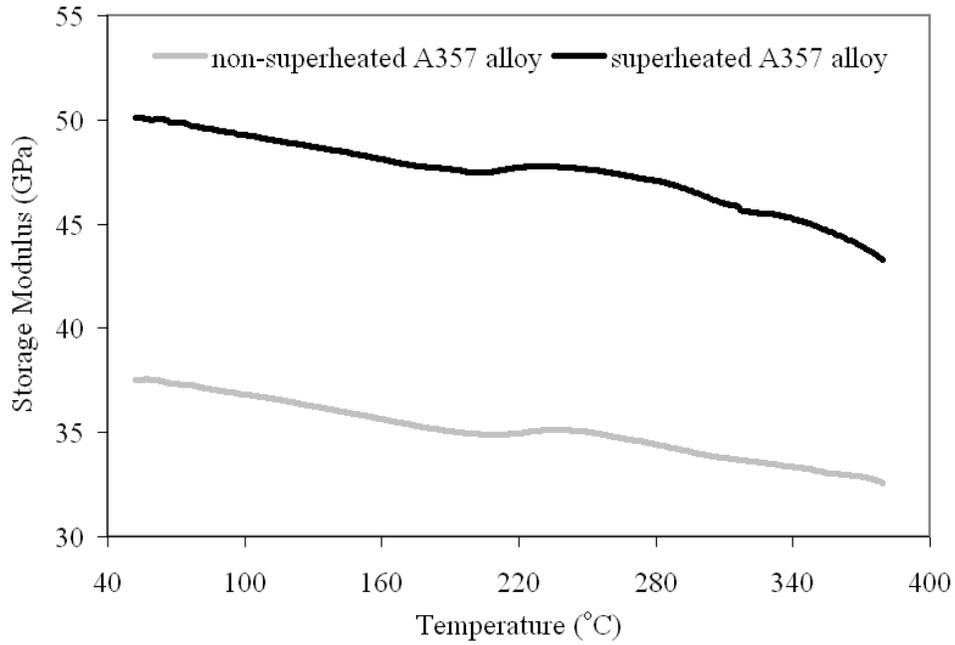

Fig. 3. The relationship of storage modulus and elevated temperatures of non-superheated and superheated A357 alloys.

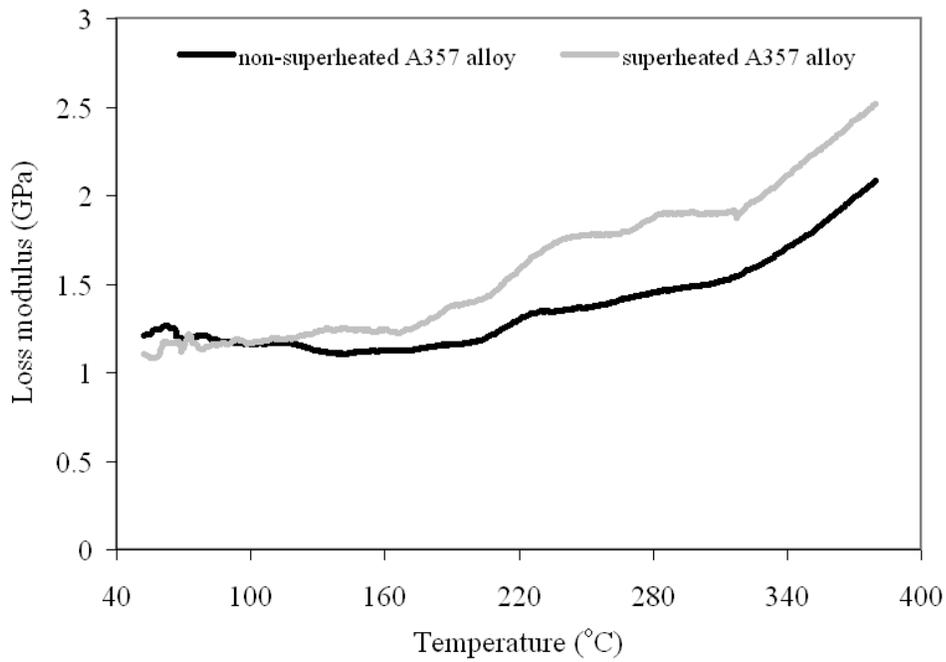

Fig. 4. The relationship of loss modulus and elevated temperatures of non-superheated and superheated A357 alloys.

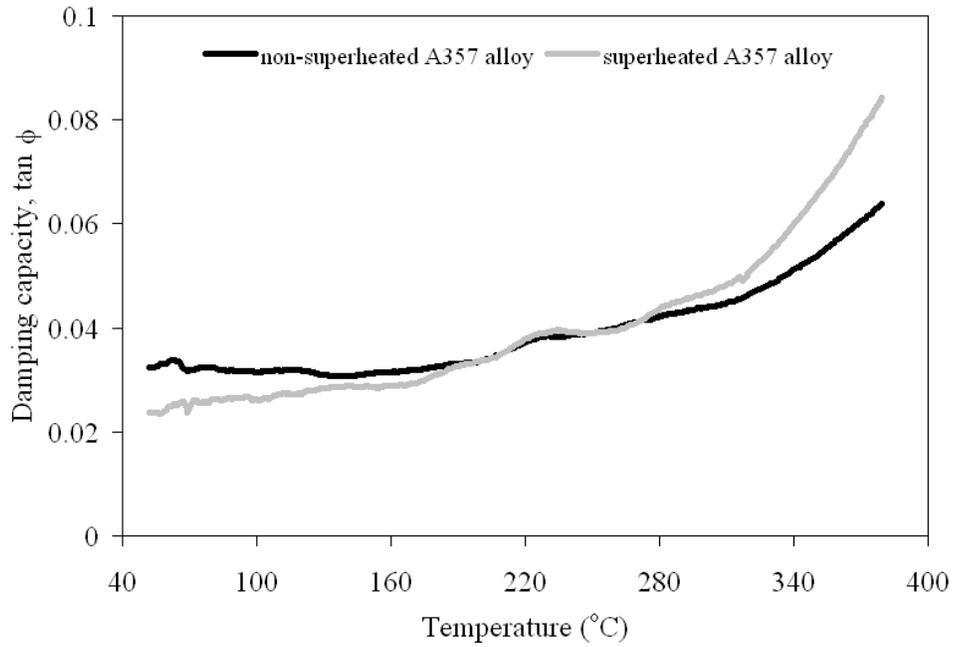

Fig. 5. The relationship of damping capacity and elevated temperatures of non-superheated and superheated A357 alloys.

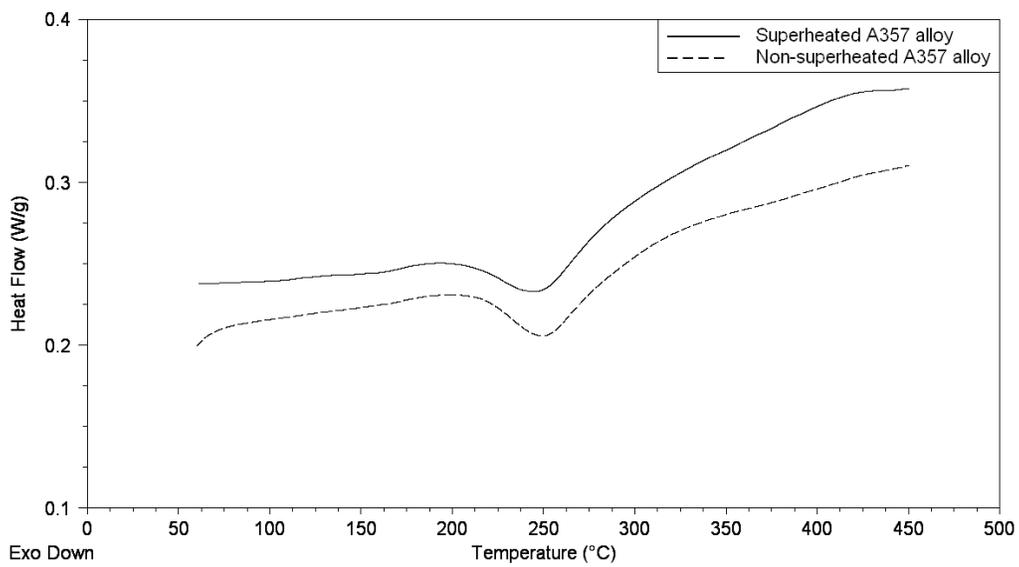

Fig. 6. The DSC thermographs of non-superheated and superheated A357 alloys.

At low temperatures, dislocation can only drag the weak pinning points such as some solute atoms and vacancies to move and then dissipating energies. At elevated temperatures, the stress for break-away from the weak pinning points is decreased because the process is thermally activated [24]. Therefore, at above certain temperatures, dislocation would move faster and then break-away dramatically from the weak pinning point. It would then become a long and comparatively long dislocation in the condition of hard pinning such as network node of dislocation and second phase, the energy dissipated by dislocation motion would not increase and thus the damping capacity values may decrease slightly. Consequently, a damping peak is caused. Afterwards, the damping capacity may increase again due to the appropriate damping mechanism at elevated temperatures.

The interface between particles and matrix is a fundamental parameter of all micro heterogeneous materials especially the properties of these materials derive from, and depend upon the transfer of load from the matrix to the reinforcing second phase. Consequently, in the past, important efforts have been invested on the one hand in characterizing the interfaces and on the other hand in establishing the role of the mismatch between the coefficients of thermal expansion (CTE) of particles and matrix on the mechanical behavior of the composite [25]. However, Zhou *et al.* [14] reported that the study of Al-Si alloys as a function of temperature has shown that an important contribution to damping is linked with dislocation movements induced by the thermal stresses due to the CTE mismatch between the aluminum matrix and silicon particles.

In present research, Fig. 5 shows one temperature peak at about 235ºC before damping capacity of non-superheated and superheated A357 alloys starts to deviate at above 277ºC. It is indicated that the temperature peak is attributed to the dislocation vibration within the Si particles. A same temperature peak has been observed at relatively lower temperatures in cast iron due to the dislocation vibration within the graphite inclusions [26]. So, it can be proposed that dislocation damping is the mechanism for non-superheated and superheated A357 alloys at lower temperatures (50 to 277ºC).

3.3.2. Grain boundary damping

Grain boundary sliding is another mechanism giving rise to damping behavior. Grain boundary damping which associated with grain boundary relaxation, anelasticity or viscosity in polycrystalline metals has been studied by Kê [27] and Zener [28]. In polycrystalline metals, the grain boundaries display viscous like properties. The viscous flow at grain boundaries will convert mechanical energy produced under cyclic shear stress into thermal energy as a result of internal friction at grain boundaries. The thermal energy will then be dissipated by the conductivity of the metal and heat exchange with the surroundings. The energy absorbed at grain boundaries is dependent on the magnitude of the shear stress and the grain boundary area per unit volume *i.e.* grain size.

Kê [27] reported that a polycrystalline aluminum showed higher damping compared to single crystal aluminum. The difference in grain boundary damping between the polycrystalline aluminum and the single crystal aluminum became manifest when the testing temperature exceeded 200ºC. In present study, a significant difference between the damping capacity of non-superheated and superheated A357 alloys at above 277ºC is illustrated in Fig. 5. Therefore, grain boundary damping is the most reliable damping mechanism for non-superheated and superheated A357 alloys at the elevated temperatures (278 to 380ºC).

## 4. Conclusions

Refinement of eutectic Si particles and smaller α-Al dendrites via superheat treatment has been contributed to the enhancement in storage modulus and damping capacity (at elevated temperatures) in superheated A357 alloy. Generally, the damping mechanisms in non-superheated and superheated A357 alloys ascribed to be a combination of dislocation damping at lower temperatures (50 to 277ºC) and grain boundary damping at elevated temperatures (278 to 380ºC).

## Acknowledgement

The authors are grateful to UniMAP for the financial support under Short Term Grant (Acc. no: 9003-00021).